\documentclass[
  journal=largetwo,
  manuscript=article-type,
  year=2020,
  volume=37,
]{cup-journal}

\usepackage{amsmath}
\usepackage{amssymb}
\usepackage[nopatch]{microtype}
\usepackage{booktabs}
\usepackage{hyperref}
\usepackage{gensymb}
\usepackage{aas-macros}
\title{A PINK update: Improvements to the CELEBI fast radio burst data reduction and analysis pipeline}

\author{M.~Glowacki}
\affiliation{Institute for Astronomy, University of Edinburgh, Royal Observatory, Edinburgh, EH9 3HJ, United Kingdom}
\alsoaffiliation{International Centre for Radio Astronomy Research, Curtin University, Bentley, WA 6102, Australia}
\alsoaffiliation{Inter-University Institute for Data Intensive Astronomy, Department of Astronomy, University of Cape Town, Cape Town, South Africa}
\email[M. Glowacki]{marcin.glowacki@roe.ac.uk}

\author{T.~Dial}
\affiliation{Centre for Astrophysics and Supercomputing, Swinburne University of Technology, Hawthorn, VIC, 3122, Australia}

\author{A. Bera}
\affiliation{International Centre for Radio Astronomy Research, Curtin University, Bentley, WA 6102, Australia}
\alsoaffiliation{ASTRON, the Netherlands Institute for Radio Astronomy, Oude Hoogeveensedijk 4,7991 PD Dwingeloo, The Netherlands}

\author{A.~T.~Deller}
\affiliation{Centre for Astrophysics and Supercomputing, Swinburne University of Technology, Hawthorn, VIC, 3122, Australia}
\alsoaffiliation{ARC Centre of Excellence for Gravitational Wave Discovery (OzGrav), Post Office Box 218, Hawthorn, VIC 3122, Australia}

\author{K.~Gourdji}
\affiliation{Australia Telescope National Facility, CSIRO, Space and Astronomy, PO Box 76, Epping, NSW 1710, Australia}

\author{A.~Jaini}
\affiliation{Centre for Astrophysics and Supercomputing, Swinburne University of Technology, Hawthorn, VIC, 3122, Australia}

\author{D.~Scott}
\affiliation{International Centre for Radio Astronomy Research, Curtin University, Bentley, WA, Australia}

\author{Y.~Wang}
\affiliation{Centre for Astrophysics and Supercomputing, Swinburne University of Technology, Hawthorn, VIC, 3122, Australia}
\alsoaffiliation{ARC Centre of Excellence for Gravitational Wave Discovery (OzGrav), Post Office Box 218, Hawthorn, VIC 3122, Australia}

\author{K.~Desnos}
\affiliation{Univ Rennes, INSA Rennes, CNRS, IETR - UMR 6164}

\author{A.~C.~Gordon}
\affiliation{Center for Interdisciplinary Exploration and Research in Astrophysics (CIERA) and Department of Physics and Astronomy, Northwestern University, Evanston, IL 60208, USA}

\author{R.~L.~Davies}
\affiliation{Centre for Astrophysics and Supercomputing, Swinburne University of Technology, Hawthorn, VIC, 3122, Australia}

\author{R.~M.~Shannon}
\affiliation{Centre for Astrophysics and Supercomputing, Swinburne University of Technology, Hawthorn, VIC, 3122, Australia}

 %% First letter not capped

\begin{document}

\begin{abstract}

Fast radio bursts (FRBs) which are well localised ($<1$") to their host galaxy are tools for studying cosmology and the intergalactic medium. Furthermore, high-time resolution datasets of their polarisation properties can enable testing of the numerous models on their potential progenitors. To that end, the CELEBI (CRAFT Effortless Localisation and Enhanced Burst Inspection) pipeline was conceived to enable data reduction from raw antenna voltages to detect fast radio transient events, localise them to sub-arcsecond precision, and produce polarimetric data at time resolutions as fine as 3~ns. Here we present a slew of updates to the CELEBI pipeline. Improvements to the astrometry correction for FRB localisations have aided our ability to determine what part of a galaxy more nearby FRBs have occurred in, which can have its own implication on the progenitor. We also have implemented time and frequency gating on detected fast transients to enable a boost to signal-to-noise, particularly useful for high dispersion measure or faint fast radio transients. We give examples of our improvements to the localisation, including for the currently `hostless' FRB\,20251019A. The polarisation calibration process has been overhauled, resulting in much more accurate measurements of derived polarisation fraction and rotation measures. Furthermore, we now have incorporated tools for structure-maximisation of the dispersion measure of fast radio transients, a software container which enables the installation of CELEBI on other machines, and improved the pipeline efficiency. Together these updates (named `Polarisation and astrometry Improvements for New Knowledge', or PINK) greatly improve our ability to keep up with the expected detection rate from the CRAFT COherent (CRACO) upgrade to the real-time fast transient detection system of the Australian SKA Pathfinder. 
\end{abstract}

\section{Introduction}

Fast radio bursts (FRBs) are bright pulses of radio emission occurring on timescales of milliseconds or less, first discovered in \cite{Lorimer2007}. FRBs have proven to be excellent probes of the hot ionised gas along their line of sight \citep{Macquart2020}, as well as for cosmological studies \citep[see review by][]{Glowacki2026}. While thousands of FRBs have been published, the majority have not been localised to their host galaxies, with most exceptions coming from repeating FRBs which make up a minority of the population \citep{ChimeFrbCollaboration2026}. The smaller fraction of localised FRBs is tied to the currently unknown origin of FRBs, although many theories exist \citep[see review by][]{Cordes2019}, and it is possible there is more than one progenitor to the observed population. The localisation of FRBs via their radio signals to sub-arcsecond precision is hence a necessity to help distinguish between these competing theories, and enable studies of the host galaxies \citep{Gordon2023}. 

The Commensal Real-time ASKAP Fast Transients survey \citep[CRAFT;][]{Macquart2010,Bannister2017,Shannon2024} with the Australian SKA Pathfinder \citep[ASKAP;][]{Deboer2009,Hotan2021} telescope has been a leader in FRB localisations to sub-arcsecond precision, by post-processing dumped voltage data %(previously spanning 3.1~s, now extended to 12.4~s) 
triggered by real-time FRB detections with corresponding bandpass and polarisation calibrator voltage data. Previous localisations of CRAFT FRBs have included one of the highest redshift FRB hosts to date at $z > 1$ \citep{Ryder2023} and the first commensal detection of an FRB with the neutral hydrogen gas in its host galaxy \citep{Glowacki2023}. The ability to detect and localise long period radio transients \citep[e.g.][]{Wang2025b} and signals from satellites \citep{James2025} are examples of other discoveries through the CRAFT survey. 

While the first several localisations by CRAFT involved processing the voltage data by hand \citep[e.g.][]{Bannister2019,Day2020}, this approach was not viable long-term, especially as the rate of detected and localised FRBs is to be significantly boosted (potentially to hundreds a year) with the new CRAFT COherent (CRACO) upgrade \citep{Wang2025}. This prompted the creation of an automated pipeline, realised as the CRAFT Effortless Localisation and Enhanced Burst Inspection pipeline \citep[CELEBI;][]{Scott2023}. In addition to producing sub-arcsecond precision positions of CRAFT FRBs, CELEBI produces polarimetric data at time resolutions as fine as 3~ns of FRB events \citep{Scott2025}. While FRBs detected through CRACO can be localised to a few arcsecond precision using visibility data, CELEBI is able to produce sub-arcsecond localisations, through direct imaging of the FRB in the voltage data and astrometric correction via radio continuum catalogues - namely the Rapid ASKAP Continuum Survey \citep[RACS;][]{McConnell2020,Hale2021}. %Since the original publication on CELEBI clear areas had been identified for improvement. These include the accuracy of the polarisation calibration with ASKAP datasets, and the accuracy and precision of fast radio transient localisation to best enable further studies, such as whether FRBs are associated with spiral arms \citep[Deller et al., in prep.,][]{Gordon2025}. Additionally, additional tools for accurate determination of the dispersion measure (DM) were identified for integration, and upgrades to the pipeline efficiency were necessary to ensure post-processing could match CRACO detection rates.   

%In the process of building CELEBI, various improvements have been made to the FRB localisation procedure. %, and the beamforming part of the pipeline which produces high-time resolution, polarisation calibrated datasets. 
%Therefore, in using the same original voltage data for previously published CRAFT FRBs, CELEBI is able to produce a smaller localisation error region, and optionally  

Here we present a new update to CELEBI, described as the `Polarisation and astrometry Improvements for New Knowledge', or PINK update. 
%Here we present updated localisations on previously published CRAFT FRBs, and introduce two specific improvements made to CELEBI for localisation:, and use of deeper ASKAP hardware correlator data for offset calculations. 
In Section~\ref{sec:localisation}, we describe improvements to our localisation of FRBs and other radio transients, including rotation of the localisation ellipse to the beam axis of the raw voltage data, and a description of a near-field imaging mode. We also introduce a new capability to `gate' FRB signals in the frequency and time domain, which boosts the signal-to-noise (S/N) and hence localisation precision, and present the effect of this method on the localisation of FRB\,20251019A, a high dispersion measure, low signal-to-noise FRB which is apparently hostless. In Section~\ref{sec:polarisation calibration}, the various improvements to our polarisation calibration workflow are described. Section~\ref{sec:tools} presents an updated version of previous work into the structure maximised value of the dispersion measure of FRBs \citep{Sutinjo2023}, improvements to the efficiency of the pipeline, and a description of the containerisation of CELEBI to enable easy installation on alternate computing systems. We summarise these improvements in Section~\ref{sec:conclusion}.

\section{Improvements to localisation}\label{sec:localisation}

We first briefly summarise CELEBI's previous functionality for localisation. Following flagging and calibration of ASKAP voltage data, a `finder bin' image is made using an optimal time-gated window (typically 10 milliseconds as a default) around the FRB event, in order to isolate the FRB emission entirely and maximise the signal to noise (S/N) in the image of the FRB \citep[fig. 5 of][]{Scott2023}.
%, CELEBI is an automated offline software pipeline that extends previous software of the CRAFT survey team to process ASKAP voltages in order to produce sub-acrsecond precision localisations of FRB events, alongside polarimetric data at 3~ns time resolution. 
Imaging the FRB alone is not sufficient to obtain a reliable FRB position, as there is an inherent offset of the ASKAP pointing that needs to be accounted for. To address this, from the remaining flagged and calibrated voltage data, a larger image of the field is created. As the FRB can be seen in multiple (overlapping) ASKAP beams, the phase centre of the field image is set to be the centre of the ASKAP beam with the highest S/N for the initial FRB detection. By cross-matching each point source identified in the image with a radio continuum catalogue through {\sc Astroquery} \citep{Ginsburg2019}, the offset is calculated between our measured source position and the catalogue position \citep[Section 3.7.4;][]{Scott2023}. A weighted mean of these offsets, multiplied by an empirical scaling factor, as in \citet{Day2020}, is used and added to the FRB position from the finder bin image to get a final position.

Previously the ASKAP voltage data used for FRB localisation spanned 3.1~s in a `4-bit' mode (in reality 8-bit complex numbers; 4 bit real, 4 bit imaginary). As of May 2024 the default voltage data form became `1-bit', allowing for voltage data dumps to extend by a factor of four to 12.4~s, as demonstrated by \citet{Dial2026}. %(Dial et al., in prep.). 
CELEBI can now process data from either mode for both fast radio transient localisation and producing HTR (high time resolution) datasets.

\subsection{Improved RACS astrometry and catalogue matching}\label{sec:racs}

Our offset correction that had been used in \citet{Scott2023} was based upon the Rapid ASKAP Continuum Survey \citep[RACS;][]{McConnell2020,Hale2021}, in particular the RACS-Low1 release (i.e. the catalogue derived from the lowest frequency ASKAP band). However, the RACS catalogue had not been astrometry corrected to account for any residual time- or direction-dependent errors between field calibrator and target scans. No phase calibration had been applied to RACS prior to imaging and self-calibration steps, and hence such errors could propagate through to the final RACS catalogue positions. As a result, systematic astrometric offsets were present in the RACS source positions and therefore also affected the field-offset calculations performed by CELEBI.

An astrometry correction for the RACS-Low1 (and later RACS-Low3, a third iteration of the full Southern sky) catalogue was derived through crossmatching with the Wide-field Infrared Survey Explorer \citep[WISE;][]{Wright2010} catalogue. Full details of the methodology are described by \cite{Jaini2025}, who demonstrated that systematic positional offsets of approximately $0.26''$--$1.24''$ could be present across the sky prior to correction. After applying the corrections, the typical positional accuracy of individual RACS sources improved to $\sim0.3''$ over most of the sky. These corrections can be significant enough to shift the localisation ellipse of distant ($z > 0.5$) FRBs onto a separate host galaxy candidate, or for nearby FRBs, shift the position on or off spiral arms. We note that localisations since and including those published by \cite{Shannon2024} include astrometry corrections for RACS.

More recently, the same correction methodology has been extended to the higher-frequency RACS catalogues \citep[RACS-Mid1 and RACS-High1;][]{Jaini2026}. %{Jaini et al., in prep}). 
These catalogues provide more appropriate reference frames for FRBs detected in the corresponding ASKAP observing bands, since they minimise systematic differences arising from frequency-dependent source structure or resolution. The CELEBI pipeline has therefore been updated to support these catalogues alongside RACS-Low. Currently, when RACS does not have sufficient coverage for field corrections (such as at low Galactic latitudes, or at high ($>30^{\circ}$) declinations where the performance of ASKAP is significantly reduced), the user is able to use a different radio continuum source catalogue \citep[e.g. VLASS;][]{Lacy2020} as an option when running the pipeline - a feature not previously implemented in CELEBI.

To support these improvements, we redesigned the catalogue lookup module used to identify reference sources in each FRB field. Earlier versions queried only the RACS-Low catalogue through the CASDA TAP service. The updated system now supports RACS-Low1, RACS-Mid1, RACS-High1, and VLASS catalogues. While RACS-Low1 continues to use unified catalogues even after astrometry corrections, the mid- and high-frequency epochs are currently distributed as beam-level source lists. CELEBI therefore dynamically assembles the relevant catalogues by combining sources from the beams closest to the target position, allowing the pipeline to construct an appropriate set of reference sources for each field while maintaining consistency with the survey data products.

We also introduced an improved source-selection procedure designed to identify reliable, compact reference sources for astrometric offset calculations. Rather than simply discarding sources with multiple matches, the new method applies three filtering criteria simultaneously: (i) a compactness requirement based on the ratio of integrated to peak flux density ($<1.5$), (ii) a S/N threshold of $>6$, and (iii) an adaptive minimum separation from neighbouring sources that increases for brighter sources. This approach reduces the impact of bright or blended sources that could bias the offset estimate while still retaining faint but isolated sources suitable for astrometric measurements. For RACS-Low, where both source and Gaussian component catalogues are available, a more sophisticated heuristic is applied to remove strongly resolved sources by identifying Gaussian components with integrated flux exceeding 20\% of the source flux and angular separations greater than $10\,\text{arcsec}$, ensuring that only compact sources are retained.

Finally, the updated CELEBI implementation includes explicit systematic uncertainty parameters for both right ascension and declination. These parameters allow the pipeline to incorporate known catalogue-level astrometric uncertainties when computing localisation offsets, providing a more realistic estimate of the final positional error budget. Together, these changes allow CELEBI to take full advantage of the improved RACS astrometry while providing a more flexible and reliable framework for identifying reference sources across multiple radio surveys.

\subsection{Localisation uncertainty along the beam axes}\label{sec:beamrot}

%\textcolor{blue}{AB is re-wording this sub-section. MG and AD need to validate it and remove all text marked in blue.}
%The first several CRAFT FRB positions were given as an error ellipse with axes in the right ascension and declination directions only. This negates to consider the orientation of the radio beam. While the effect on the position will typically be relatively minor, we expect that the true localisation ellipse would be aligned with the radio beam. Since \citet{Scott2023} the radio beam orientation has been implemented, and used for localisations reported in e.g. \cite{Shannon2024} and following CRAFT publications. 
Position errors for localised FRBs are most commonly expressed in terms of uncertainties along the axes of the celestial coordinate system in use -- typically right ascension and declination -- which is equivalent to an uncertainty ellipse aligned with those axes. However, since the `natural' basis for the interferometric localisation errors is the coordinate system defined by the major and minor axes of the synthesised beam, the `true' position uncertainty ellipse is aligned with the beam axes rather than the celestial coordinate system. We hence estimate the absolute astrometric offsets and position uncertainties along the axes of the beam.

For each reference field source, astrometric offsets projected along the beam major and minor axes are calculated using a co-ordinate transform given by
\begin{equation}
    \label{eqn:offset}
    \begin{split}
    \Delta X_{\rm major} = \Delta {\rm RA} \: \sin{({\rm BPA})} + \Delta {\rm DEC} \: \cos{({\rm BPA})} \\
    \Delta X_{\rm minor} = \Delta {\rm RA} \: \cos{({\rm BPA})} - \Delta {\rm DEC} \: \sin{({\rm BPA})},
    \end{split}
\end{equation}
where $\Delta {\rm RA}$, $\Delta {\rm DEC}$ are the differences between the source co-ordinates in the field image and those in the reference catalogue, and BPA is the position angle (in degrees, although radians can also be used here and equations 2--4) of the (elliptical) synthesised beam associated with the field image. Uncertainties on the astrometric offset for each reference source are then estimated using the relation  
\begin{equation}
    \label{eqn:offset_error}
    \begin{split}
    \delta X_{\rm major}^2 = [d_{\rm maj} \: \cos{(\Delta{\rm PA})}]^2 + [d_{\rm min} \: \sin{(\Delta{\rm PA})}]^2 + \sigma_{\rm ref}^2 \\
    \delta X_{\rm minor}^2 = [d_{\rm maj} \: \sin{(\Delta{\rm PA})}]^2 + [d_{\rm min} \: \cos{(\Delta{\rm PA})}]^2 + \sigma_{\rm ref}^2, \\
    \end{split}
\end{equation}
where $d_{\rm maj}, d_{\rm min}$ represent the major and minor axes of the best-fit elliptical Gaussian describing the source and $\Delta{\rm PA}$ is the difference between its position angle and BPA; $\sigma_{\rm ref}$ represents the position uncertainty of the source in the reference catalogue (RACS) which is 
treated as a circularly symmetric quantity, and is significantly sub-dominant to the field image uncertainties, as RACS is considerably deeper than the 12\,s field image.
%approximately symmetric. \textcolor{blue}{[AB: This is what we are currently assuming in the offset correction code. Needs validation by Adam. ADAM: Akhil would know exactly what the per source uncertainties are in RACS, but in any case they should always be far subdominant to the field image uncertainties, since we're comparing a 12s image to a 15 minute one. So I would think we could say something like "which is treated as a circularly symmetric quantity, and is significantly sub-dominant to the field image uncertainties, as RACS is considerably deeper than the 12\,s field image."]} 
The inverse-variance-weighted mean of the offsets, calculated separately along the major and minor axes, represents the best estimate of the net astrometric offset and the associated uncertainties. 

The error on the FRB position has three independent contributors -- the statistical error on the FRB position in the image, statistical error on the offset from the reference catalogue, and systematic error associated with the reference catalogue astrometry -- which are added in quadrature separately along the beam major and minor axes. The statistical error on FRB position in the image, along the synthesised beam axes, are given by
\begin{equation}
    \label{eqn:stat_error}
    \begin{split}
    \delta Stat_{\rm major}^2 = \frac{\sqrt{[D_{\rm maj} \: \cos{(\delta{\rm PA})}]^2 + [D_{\rm min} \: \sin{(\delta{\rm PA})}]^2}}{2.35 \: {\rm SNR}}  \\
    \delta Stat_{\rm minor}^2 = \frac{\sqrt{[D_{\rm maj} \: \sin{(\delta{\rm PA})}]^2 + [D_{\rm min} \: \cos{(\delta{\rm PA})}]^2}}{2.35 \: {\rm SNR}},  \\
    \end{split}
\end{equation}
where $D_{\rm maj}, D_{\rm min}$ represent the major and minor axes of the best-fit elliptical Gaussian describing the FRB in the image and $\delta{\rm PA}$ is the difference between its position angle and BPA; SNR is the image-plane signal-to-noise ratio of the FRB. The statistical errors on the offset from the reference catalogue, along the major and minor axes of the synthesised beam, are given by
\begin{equation*}
    1.79 \: (\delta X_{\rm major}, \delta X_{\rm minor})
\end{equation*}
where the systematic scale factor of 1.79 was empirically estimated by \citet{Day2021} and incorporates systematic contributions to positional offsets for each source such as the mismatch in angular resolution and frequency between the two images being compared. The systematic astrometric uncertainty of the reference catalogue is adopted from \citet{Jaini2025}, corresponding to $0.30''$ over most of the RACS sky and degrading slightly to $0.40''$ within the Galactic plane.
% \textcolor{blue}{[AB: One of Akhil's papers! Leaving for AJ to complete the sentence.]} and projected onto the beam axes using a simple rotation of co-ordinates.

\begin{figure*}
\centering
\includegraphics[width=0.33\textwidth]{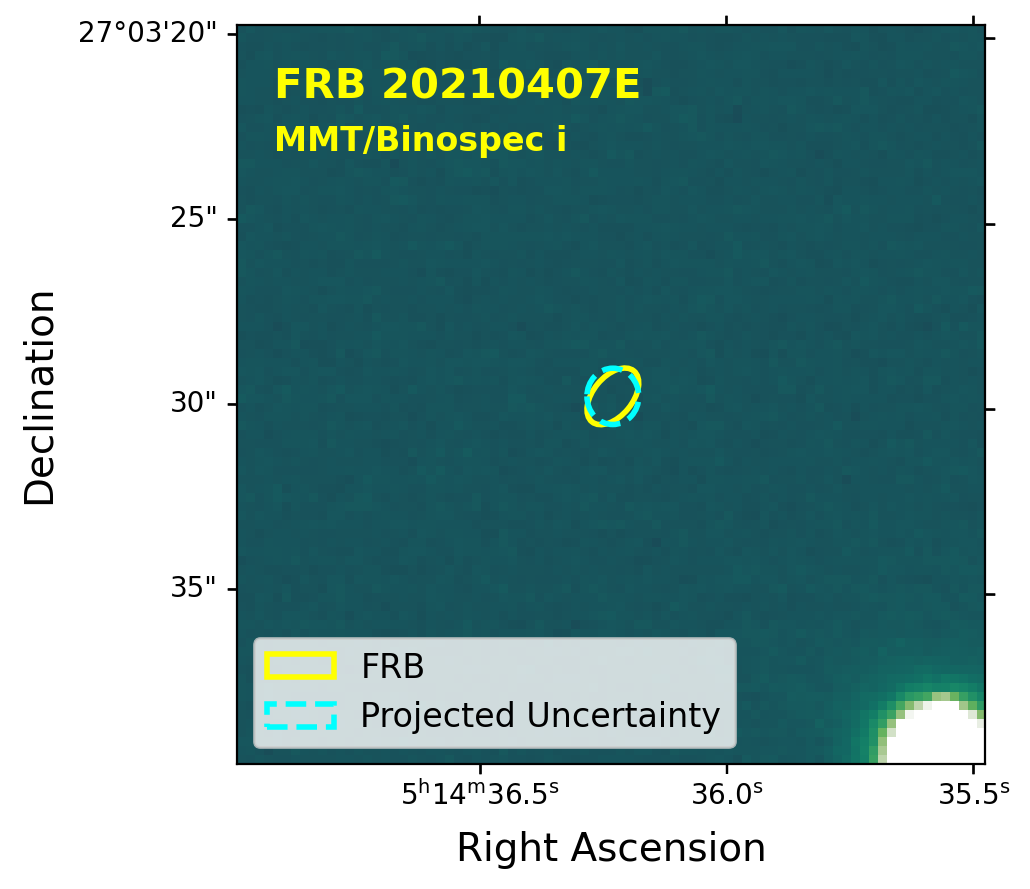}
\includegraphics[width=0.33\textwidth]{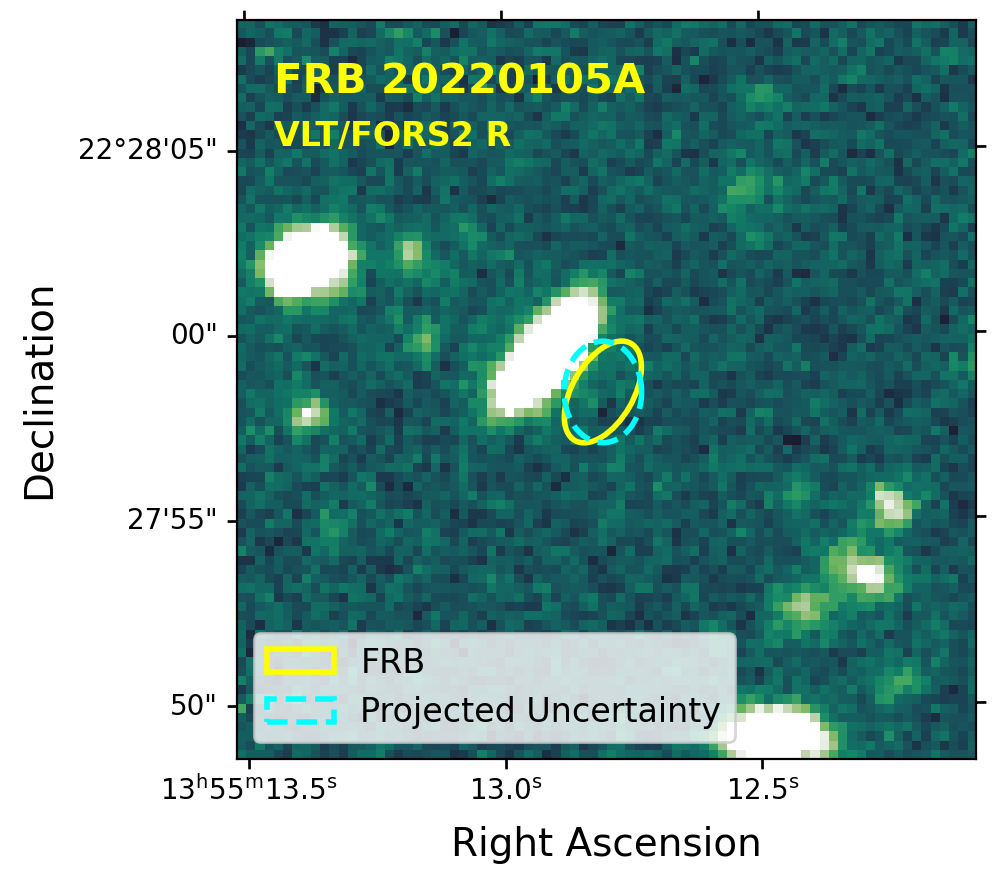}
\includegraphics[width=0.33\textwidth]{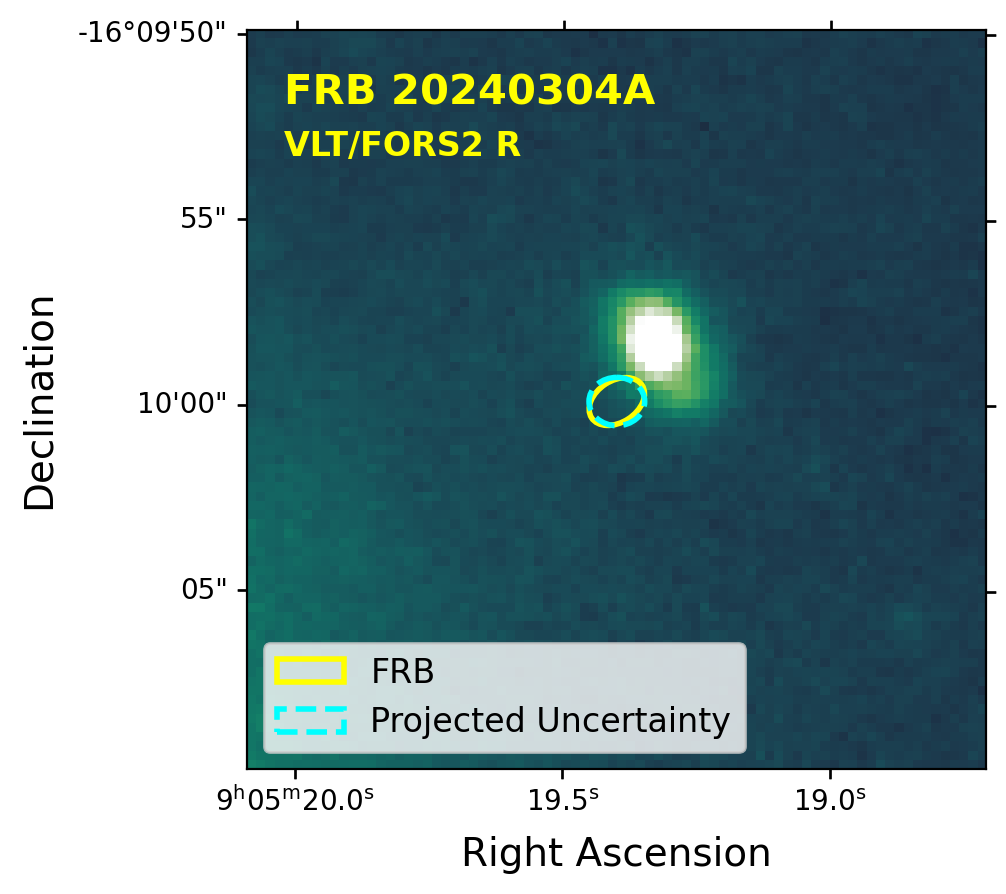}
\caption{The projected error (dashed cyan) and true error (solid yellow) ellipses for three CRAFT FRBs. The background optical imaging span 20 arcseconds across and were originally presented in \citealt{Shannon2024}.}
\label{fig:img_examples}
\end{figure*}

The uncertainties on the localisation of a given FRB are quoted in terms of an error ellipse with axes ($\delta_{\rm major}, \delta_{\rm minor}$), oriented along the synthesised beam in the corresponding field image with a position angle defined by BPA. The projected uncertainties along RA and DEC axes are given by
\begin{equation}
    \label{eqn:pos_error}
    \begin{split}
    \delta {\rm RA} = \sqrt{[\delta_{\rm major} \sin{({\rm BPA})}]^2 + [\delta_{\rm minor} \cos{({\rm BPA})}]^2} \\
    \delta {\rm DEC} = \sqrt{[\delta_{\rm major} \cos{({\rm BPA})}]^2 + [\delta_{\rm minor} \sin{({\rm BPA})}]^2}. \\
    \end{split}
\end{equation}
However, as shown in Figure~\ref{fig:img_examples}, the error-ellipse implied by these projected errors (from left to right 0.7"~x~0.76", 1.045"~x~1.369", and 0.75"~x~0.65"; represented by dashed cyan ellipses) does not represent the true error ellipse (ellipse major axis, minor axis, and PA of [0.89", 0.53", -39.6$^{\circ}$], [1.513", 0.823", -30.45$^{\circ}$], [0.81", 0.56", -57.04$^{\circ}$]; denoted in solid yellow). For any analysis sensitive to the FRB position accuracy (e.g., the location relative to spiral sub-structure; \citealt{Gordon2025}), using the true error-ellipse is recommended\footnote{We note that the astrometric uncertainty of optical imaging is also important for analysis of the location of the FRB with respect to e.g. spiral arms in its host galaxy, but this is separate to any FRB localisation made by CELEBI}.%when performing PATH analysis}.}. % FRB location analysis 

\subsection{Matched filter imaging}\label{sec:matched}

\label{sec:mfimage}

% Description here. We initially use the full frequency range, sans RFI flagging, and assume 20~ms intervals, for producing finder bin images. FRBs do not necessarily occupy the full 336~MHz bandwidth of ASKAP, and can be substantially narrower than 20 ms in span (quote number range from HTR paper). 

% We use HTR outputs to determine the frequency and time range, and feed this into the localisation step.

% Example or two of improvement to S/N, and change in localisation. Quantify typical improvement here.

The previous iteration of CELEBI \citep{Scott2023} made use of large boxcar filters centred on the rough position of the FRB in the voltage buffer to image and produce a localisation with (sub)arcsecond precision. The boxcar would typically be of O(10) ms wide, which for most CRAFT FRBs, having widths of O(100) $\mu$s to O(1) ms \citep{Scott2025}, would introduce a large amount of noise. Additionally, when adding the visibilities together, further noise would be introduced if the FRB possessed complex morphology in time and/or frequency (e.g. scintillation or band-limited emission). This becomes an issue when faint FRBs are detected, especially with the recent CRACO upgrade which has increased the rate of faint FRB detections. Hence, a new process called matched filter imaging has been implemented into CELEBI.

The new imaging pipeline makes use of the high time resolution (HTR) Stokes $I$ power dynamic spectra to build an optimal time and frequency matched filter, which when applied will minimise the noise added when imaging the FRB. To build the matched filter, the FRB needs to be located in the HTR data, enabling the signal--to--noise ratio as a function of time and frequency to be measured. Producing this HTR data via beamforming requires an initial estimate of the FRB location, which can be provided from the CRACO detection if available, or from a default boxcar filter imaging otherwise. Once generated, the Stokes $I$ dynamic spectrum is averaged in frequency. Then, the on-pulse window of the burst is estimated using a cutoff threshold which is some fraction of the peak sample. Time-dependent and frequency-dependent weights are then derived using the on-pulse window:

\begin{equation}
    \begin{split}
        &w_{t} = \frac{1}{n}\sum_{f}^{n} I(t,f)\\
        &w_{f} = \frac{1}{m}\sum_{t}^{m} I(t,f)
    \end{split}
    \label{eq:weights}
\end{equation}

\noindent where $w_{t}$ and $w_{f}$ are time dependent and frequency dependent weights respectively, $m$ and $n$ are the number of time bins and frequency channels respectively. We note that time-dependent weights are always used for matched filter imaging, while frequency weighting is turned off by default. Frequency weighting is only switched on when the final FRB position is signal--to--noise ratio limited and the FRB has significant frequency structure (generally, is only bright in a small fraction of the bandwidth), such that applying frequency weighting leads to an appreciable gain in detection significance. Furthermore, since the HTR data products are normalised to unity variance in every frequency channel, the frequency weighting does not take into account the instrumental bandpass response.
An off-pulse window is also taken for RFI subtraction. The off-pulse window is split into two equal parts and placed on either side of the on-pulse window. A guard window is then placed between the on-pulse and off-pulse windows to avoid potential signal leakage into the RFI bins. The typical width of the off-pulse window is of O(10) ms whilst the width of the guard window is of O(1) ms. This is to ensure most of the RFI being subtracted is constant in time \citep{Scott2023}. The RFI bins, guard bins and on-pulse bins are combined into a single matched filter binconfig file and passed to DiFX to correlate and obtain visibility data. 

The time and frequency weights are applied to the visibilities separately, with the time weighting applied immediately as the data were time-scrunched (averaged to a single time bin), while the frequency weights are applied during the gridding stage when imaging the FRB. Since the FRB model at the imaging stage uses a frequency independent model for the FRB emission, the frequency weights derived in Equation~\ref{eq:weights} are used to calculate both an amplitude correction to the visibilities (required to make the FRB appear spectrally flat; \citealp{rau2011multi}), and a modification to the corresponding visibility weights. Essentially, this upweights frequency ranges where the FRB is bright (even after the visibility amplitude has been scaled down) and downweights frequency ranges where it is faint (even after the visibility amplitude is scaled up). The time scrunching process generates a single visibility per frequency (and per baseline, but the baseline index is not shown here as all baselines receive the same scaling) as follows: 

\begin{equation}
    V(f) = \frac{1/w_{f}\sum^{n}(V(t,f) - V(f)_{\mathrm{RFI}})w_{d}(t,f)w_{t}}{\sum^{n}w_{d}(t,f)w_{t}}.
\end{equation}

\noindent Here, $w_{d}(t,f)$ are the initial visibility weights (which are derived during correlation and account for the number of data samples that have contributed to each visibility, as a function of time and frequency) and $V(f)_{\mathrm{RFI}}$ are the time averaged RFI visibilities. $1/w_{f}$ is the appropriate amplitude scaling factor to scale the FRB amplitude to a constant value as a function of frequency. The frequency-dependent visibility weights that are saved to the visibility dataset and applied in the imaging process are re-calculated to account for this frequency dependent amplitude correction as well as the summation over time:

\begin{equation}
    w_{d}(f) = w_{f}^2\sum^{n}w_{d}(t,f)w_{t},
\end{equation}

\noindent which allows for the bright regions of FRB emission (where the noise is now low, due to the amplitude scaling) to be upweighted during imaging, and the faint regions (where the noise is high) to be downweighted. The result of this process is a single match-filtered visibility time bin that is then processed in the same manner described in Section 3.7.2 of \citet{Scott2023} to image the FRB and produce a more refined localisation. In turn, the refined localisation can then be used to ensure an optimally coherent addition of the signals from every antenna -- maximising the signal-to-noise ratio -- in new HTR outputs. 

%\textbf{Example (or two?) here for the improvement to S/N for FRBs with the matched filter approach - image included - and also the resulting improvement to localisation as a result (ACG to do using 251019 example)}

\subsubsection{A matched filter example case: FRB\,20251019A}

The use of the matched filter imaging approach is particularly important for lower S/N FRBs only detected by the CRACO system, as well as high-DM FRBs which are likely at high redshift. Distant FRBs require as high a precision as possible for accurate host association. An example of such an object is FRB\,20251019A, which was detected with ASKAP/CRACO at a DM of 1277~pc\,cm$^{-3}$ and localised to sub-arcsecond precision with a reported S/N of 9.9. The localisation uncertainty ellipse originally from CELEBI using 5\,ms of data had a major/minor axis size of 0.727/0.575", while implementation of matched filter imaging (time weighting only; the FRB did not have significant frequency structure) reduced this to 0.594/0.507" centred on 10:08:27.454, +16:23:56.358 (reduction of 18\%/12\% respectively, and an unchanged PA of -38.06$^{\circ}$). 

\begin{figure}
    \centering
    \includegraphics[width=\linewidth]{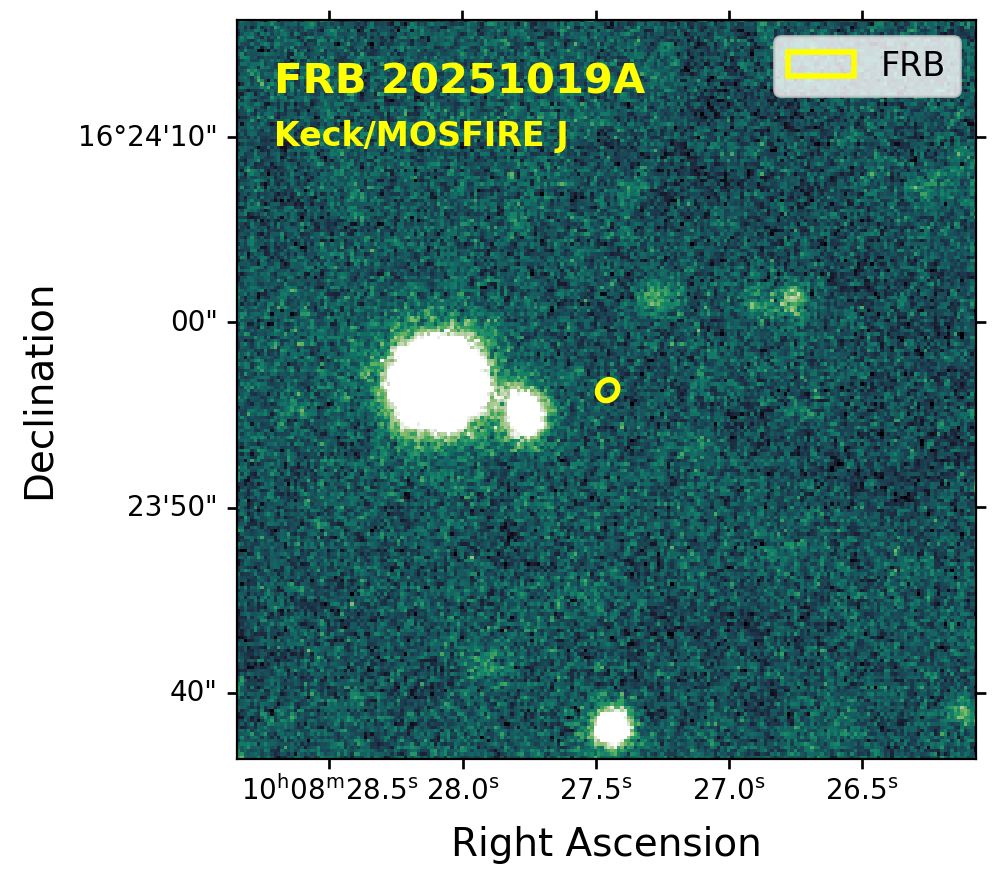}
    \caption{The $J$-band Keck/MOSFIRE image of the field of FRB\,20251019A. The 1$\sigma$ FRB localisation is overlaid as a yellow ellipse; no source is visible at the FRB position to a 3$\sigma$ limiting magnitude of $J_{\rm AB} > 24.4$.}
    \label{fig:FRB20251019_Jband}
\end{figure}

To identify the host galaxy of FRB\,20251019A, we performed a PATH \citep{path} analysis on archival $r$-band imaging from the DESI Legacy Survey DR9 \citep{Legacy}. However, this analysis revealed no viable host candidates and a large posterior that the host was unseen (P(U|x) = 0.99). We obtained Keck/MOSFIRE \citep{MOSFIRE} imaging in $J$-band for a total exposure time of 1375s on 2026 Jan 15 UT (PI Davies, Program 2025B\_W007). The data were reduced using the \texttt{POTPyRI}\footnote{\url{https://github.com/CIERA-Transients/POTPyRI}} imaging reduction pipeline. However, no host galaxy was identified at or around the FRB position to a 3$\sigma$ limiting magnitude of $J_{\rm AB} > 24.4$. We show the image and 1$\sigma$ FRB localisation in Figure~\ref{fig:FRB20251019_Jband}. Despite the elusiveness of the host of FRB\,20251019A to date, matched filter imaging enables the best possible localisation for such FRBs and hence maximises the ability to localise currently rare high-redshift ($z > 1$) FRB candidates. This object, along with other currently hostless FRBs such as FRB\,20210912A \citep{Marnoch2023}, highlights the need for deep optical imaging for the higher redshift FRB population.

\subsection{Near-field imaging}\label{sec:nearfield}
Both the correlation and beamforming functions of the CELEBI pipeline require a model of the geometric propagation delay of the signal wavefront between the ASKAP antennas. By default, this is provided by the CALC package used by the DiFX software correlator, which is a highly accurate and well tested suite of code used for astrophysical sources. However, this assumes an astrophysical source located far from the array. For near-field objects where the curvature of the signal wavefront is not negligible, a different approach to the generation of the geometric delay model is required.

For near-field objects, the DiFX delay model generation step can instead make use of the NASA SPICE toolkit\footnote{\url{https://naif.jpl.nasa.gov/naif/toolkit.html}} to produce the geometric delays used in correlation and beamforming. While DiFX has provided this functionality for some time \citep[e.g.][]{McCallum2025}, the CELEBI interface prior to CRACO detections did not initially expose the ability to demand a near-field geometric model, nor the ability to provide the Two-Line Ephemeris (TLE) information needed to generate such a model. In this updated version of CELEBI, target sources can be specified as near-field objects and ephemeris information provided in order to accurately image and/or beamform such objects. For example, when CRAFT detected a nanosecond-duration radio pulse from the Relay-2 satellite \citep{James2025}, this functionality was used to search for fainter pulses from the satellite during the recorded voltage download by accurately tracking the satellite for the duration of the 12\,s data window.

\subsection{Single polarisation processing}
Data were downloaded from the ASKAP antennas by coarse channel, antenna, and polarisation in a sequential fashion, with polarisation being the slowest variable to be downloaded. Accordingly, if a voltage download is interrupted (for instance, due to a scan change which is more likely for an FRB detected during shorter calibration observations), it is possible that only data from a single polarisation (possibly for a subset of antennas) is obtained. In the initial version of CELEBI, such single polarisation datasets could only be processed for calibration and imaging through manual intervention (by trimming the calibrator voltage dataset to exactly match the obtained FRB dataset, in terms of antennas and polarisations available), while beamforming assumed dual polarisation and was completely intractable.

In this version of CELEBI, we have generalised the application of calibration solutions in both imaging and beamforming (as well as the production of Stokes data products in the beamforming pipeline) to gracefully handle single polarisation datasets, producing a pseudo-Stokes I image and pseudo-Stokes I time series (without any of the polarisation calibration described in Section~\ref{sec:polarisation calibration} below) from the available polarisation data. This has allowed the recovery of CELEBI positions and time series data for {\bf 3} FRBs, with two newer FRBs since \citet{Scott2025} to be presented in a forthcoming ASKAP data release (Dial et al., in prep).

\subsection{Deeper field imaging for offset corrections}\label{sec:deeperimage}

There are two main components of fast radio transient localisation: the position of the imaged transient itself, and the offset correction based on the positions of detected field sources in the remaining voltage dump data, currently limited to 12 seconds of integration time. One avenue to further improve the accuracy of FRB localisation is to use a deeper image of the field from the ASKAP observation the radio transient was detected within. For instance, typical RACS pointings are $\sim$15 minutes each in length, and would yield many more field sources than what is visible in 12 seconds (a signal-to-noise boost of 8.66). This would be particularly beneficial for any transient sources detected in a relatively `quiet' part of the radio sky containing a low number of sufficiently bright radio continuum sources that can be imaged in 12 seconds than the number that would appear in a 15 minute ASKAP integration.

However, the ASKAP hardware data and the dumped voltage-based data, as well as calibration and flagging applied to each, may not be equivalent. Hence a further correction may need to be applied when using a deeper field image from the hardware data. A separate approach that could be taken is to measure the offset of field source positions between the deeper field image and the 12 second derived field image, and in turn use the deeper image for a better measurement of the offset field correction, benefiting from the larger number of sources. A proper investigation and implementation of this approach is planned in a later update to CELEBI. 

\section{Polarisation Calibration}

\label{sec:polarisation calibration}

A number of instrumental effects can alter the measured polarisation of a detected FRB in undesirable ways. Consequently, a calibrator with a known polarisation is used to model and mitigate these effects. Since the last iteration \citep{Scott2023}, major improvements have been made to polarisation models and their application, which are detailed herein.

\subsection{Polarimetric effects}

The polarisation of our radio signal is characterised by the Stokes parameters, which are expressed in terms of horizontal and vertical components $\mathrm{E_{x}}$ and $\mathrm{E_{y}}$:

\begin{align}
    \begin{split}
        & I = \mathrm{|E_{x}|^{2} + |E_{y}|^{2}}\\
        & Q = \mathrm{|E_{y}|^{2} - |E_{x}|^{2}}\\
        & U = 2\mathrm{Re(E_{x}^{*}E_{y})}\\
        & V = 2\mathrm{Im(E_{x}^{*}E_{y})},
    \end{split}
    \label{eq:Stokes}
\end{align}

\noindent where Stokes $Q$ is negated to comply with the pulsar Stokes conventions in \citet{van2010psrchive}, as the ASKAP Phased Array Feed (PAF) uses a left-handed basis \citep{Day2020}. We use a polarisation calibrator with known rest-frame polarisation fractions:

\begin{align}
    \mathbf{S_{0}}
    =
    \begin{bmatrix}
        Q_{\mathrm{o}}(f)\\
        0\\
        V_{\mathrm{o}}(f)
    \end{bmatrix}.
\end{align}

\noindent The signal undergoes Faraday rotation along the line of sight (LOS). This is modelled as a simple rotation $\chi$ about the $(Q,U)$ plane (i.e. a rotation of the linear polarisation vector): 

\begin{equation}
    \mathbf{A}
    =
    \begin{bmatrix}
        \mathrm{cos}(2\chi) & \mathrm{sin}(2\chi) & 0 \\
        -\mathrm{sin}(2\chi) & \mathrm{cos}(2\chi) & 0 \\
        0 & 0 & 1
    \end{bmatrix}.
    \label{eq:polcal_FR}
\end{equation}

\noindent Once a signal reaches the ASKAP antenna, the PAF will have an angular offset  $\psi$ from the sky coordinate system \citep{Day2020}, which may also be expressed as a rotation in the $(Q,U)$ plane:

\begin{equation}
    \mathbf{A} = 
    \begin{bmatrix}
        \mathrm{cos}(2\chi + 2\psi) & \mathrm{sin}(2\chi + 2\psi) & 0 \\
        -\mathrm{sin}(2\chi + 2\psi) & \mathrm{cos}(2\chi + 2\psi) & 0 \\
        0 & 0 & 1
    \end{bmatrix}.
\end{equation}

\noindent The electronics that separately digitise the $\mathrm{E_{x}}$ and $\mathrm{E_{y}}$ components are imperfect; as such, they introduce a slight time and phase delay ($\tau$, $\phi$) during recombination, known as polarisation leakage. This results in conversion between linear and circular polarisation, expressed as a rotation in the $(U,V)$ plane by an angle $\theta$:

\begin{equation}
\begin{split}
    \theta &= 2\pi f\tau + \phi \\
    \mathbf{B}
    &=
    \begin{bmatrix}
    1 & 0 & 0 \\
    0 & \mathrm{cos}(\theta) & \mathrm{sin}(\theta) \\
    0 & -\mathrm{sin}(\theta) & \mathrm{cos}(\theta)
    \end{bmatrix}.
    \label{eq:leakage_angle}
    \end{split}
\end{equation}

\noindent Finally, the PAF edge and corner beams are known to exhibit significant leakage, causing coupling between the $\mathrm{E_{x}}$ and $\mathrm{E_{y}}$ components. Consequently, the natural wave modes received by the antenna become slightly elliptical. The most severe cases show $\sim$5-10\% coupling. \cite{heiles2001mueller} formalises these instrumentation effects more generally. Here, we make the assumption that the phase angle of coupling between $\mathrm{E_{x}}$ and $\mathrm{E_{y}}$ (i.e. $\chi$ in \cite{heiles2001mueller}) is 90$^{\circ}$, which reduces the coupling to a simple rotation in the (Q, V) plane by the ellipticity angle $\alpha$:

\begin{equation}
    \mathbf{C}
    = 
    \begin{bmatrix}
    \mathrm{cos}(2\alpha) & 0 & \mathrm{sin}(2\alpha) \\
    0 & 1 & 0 \\
    -\mathrm{sin}(2\alpha) & 0 & \mathrm{cos}(2\alpha)
    \end{bmatrix}.
    \label{eq:ellipticity}
\end{equation}

Accounting for these effects, the measured Stokes parameters $\mathbf{S_{m}}$ become:
% These polarimetry effects have been reduced to rotations in 3D space $(Q,U,V)$, meaning the order of corrections is important. Faraday rotation occurred first, followed by the angular mis-alignment of the PAF, coupling of the $\mathrm{E_{x}}$ and $\mathrm{E_{y}}$ components and finally polarisation leakage due to imperfect electronics during digitisation. The measured Stokes parameters then become

\begin{equation}
    \mathbf{S_{m}}
    =
    \mathbf{B}
    \mathbf{C}
    \mathbf{A}
    \mathbf{S_{0}}.
    \label{eq:measured_stokes}
\end{equation}
\noindent By default, we do not fit $\alpha$ because not all FRBs are detected in corner or edge beams where this effect is significant. Hence, this parameter is a toggle in the CELEBI implementation. 

\subsection{Deriving calibration solutions}
\label{sec:derive_polcal}

Using Eq.\ref{eq:measured_stokes} we sample the full parameter space ($\mathrm{RM}$, $\psi$, $\tau$, $\phi$, $\alpha$, $L_{\mathrm{offset}}$, $V_{\mathrm{offset}}$) to derive polarisation calibration solutions. The two additional free parameters $L_{\mathrm{offset}}$ and $V_{\mathrm{offset}}$ are sampled to account for small zeroth-order deviations in the total linear and circular polarisation fractions over different observing bands:

\begin{equation}
    \begin{split}
        & L_{m} \rightarrow L_{m} + L_\mathrm{offset} \\
        & V_{m} \rightarrow V_{m} + V_\mathrm{offset}.    
    \end{split}
\end{equation}

\noindent Using Bayesian inference with the \textsc{dynesty} (Dynamic nested sampling) Sampler \citep{speagle2020dynesty} through the \textsc{Bilby} \citep{ashton2019bilby} package, we  maximise the log likelihood $\mathcal{L}_{T}$:

\begin{equation}
\begin{split}
    & \mathcal{L}_{S} = -\frac{1}{2}\sum^{N}_{i = 1}\Bigg(\ln(2\pi\sigma^{2}_{S}) + \bigg(\frac{S-S_{m}}{\sigma_{S}}\bigg)^{2} \Bigg) \\
    & \mathcal{L}_{T} = \mathcal{L}_{Q} + \mathcal{L}_{U} + \mathcal{L}_{V},
\end{split}
\end{equation}

\noindent where $\sigma_{S}$ is the error measurement in the measured Stokes parameter $S$, $S_{m}$ is the modelled Stokes parameters using Eq.\ref{eq:measured_stokes} and $N$ is the number of frequency channels.

\subsection{Applying calibration solutions}

Applying the solutions for the polarimetric effects detailed above is trivial in (Q, U, V) space. However, these solutions must be applied when constructing the Stokes dynamic spectra, which is performed multiple times for some FRBs - such as when channelising at different frequency resolutions, applying baseline corrections or performing coherent de-dispersion. A more robust, albeit more complex, method is to apply these solutions directly to the $\mathrm{E_{x}}$ and $\mathrm{E_{y}}$ voltages after coherent beamforming. 

First, the polarisation leakage corrections $\tau$ and $\phi$ are applied to the $E_{x}$ voltages in the frequency domain:

\begin{align}
    \mathrm{E_{x}}(t) = \mathcal{F}^{-1}\bigg[\mathrm{E_{x}}(f)\mathrm{e}^{-i\big(2\pi f\tau + \phi\big)}\bigg],
\end{align}

\noindent where $\mathcal{F}^{-1}$ is the inverse Fourier transform. Next, the coupling between $\mathrm{E_{x}}$ and $\mathrm{E_{y}}$ (if applicable) is corrected by rotating $\mathrm{E_{x}}$ and $\mathrm{E_{y}}$ by the ellipticity angle $\alpha$ in the time domain:

\begin{equation}
    \begin{bmatrix}
        \mathrm{E_{x}} \\
        \mathrm{E_{y}}
    \end{bmatrix}
    =
    \begin{bmatrix}
        \mathrm{cos}(\alpha) & -i\mathrm{sin}(\alpha) \\
        -i\mathrm{sin}(\alpha) & \mathrm{cos}(\alpha)
    \end{bmatrix}
    \begin{bmatrix}
        \mathrm{E_{x}} \\
        \mathrm{E_{y}}
    \end{bmatrix}.
\end{equation}

\noindent Finally, the PAF angular misalignment is corrected by rotating $\mathrm{E_{x}}$ and $\mathrm{E_{y}}$ by $\psi$:

\begin{equation}
    \begin{bmatrix}
        \mathrm{E_{x}} \\
        \mathrm{E_{y}}
    \end{bmatrix}
    =
    \begin{bmatrix}
        \mathrm{cos}(\psi) & -\mathrm{sin}(\psi) \\
        \mathrm{sin}(\psi) & \mathrm{cos}(\psi)
    \end{bmatrix}
    \begin{bmatrix}
        \mathrm{E_{x}} \\
        \mathrm{E_{y}}
    \end{bmatrix}.
\end{equation}

\noindent Note that this is an inverse rotation, since we are correcting the physical rotational offset.

\subsection{Polarisation calibrators}

Two sources are used for polarisation calibration: J0835-4510 and J1644-4559 (hereafter Vela and 1644, respectively). Similarly to \citet{Scott2023}, we model 1644 using a second order polynomial:

\begin{equation}
\begin{split}
    & L_{\mathrm{o}} = -1.31\times10^{-8}f^{2} + 1.76\times10^{-4}f - 3.18\times10^{-2} \\
    & V_{\mathrm{o}} = 3.19\times10^{-8}f^{2} -1.82\times10^{-6}f -1.2\times10^{-2},
\end{split}
\end{equation}

\noindent where $f$ is the channel frequency in MHz. While 1644 is well-behaved across the mid-high ASKAP bands ($f~>~1$ GHz), two properties hinder calibration at lower frequencies. (1) The long rotation period and limited ASKAP voltage buffer allow only a small number of pulses to be folded, limiting the signal-to-noise. (2) A large scattering timescale causes the pulse profile to broaden significantly, depolarising the burst. Thus, caution is required when using 1644 at low frequencies. 

For Vela, we provide two models. The first assumes a flat spectrum where $L_{\mathrm{o}} = 0.95$ and $V_{\mathrm{o}} = -0.05$ \citep{Scott2023}, as used in recent CRAFT publications \citep{Bera2024, Dial2025, Scott2025}. The second model uses Ultra wideband Low (UWL) observations of Vela with the Murriyang (Parkes) telescope \citep{hobbs2020ultra}. UWL observations of Vela from December 2020 to December 2024 were reduced using \textsc{PSRCHIVE} \citep{hotan2004psrchive}, including RFI flagging, band-pass calibration and de-dispersion. The Stokes dynamic spectra were then dumped from the archive files and an additional incoherent de-dispersion search was used to maximise the signal-to-noise in the Stokes I time-series profile. We integrated $\sim160$ $\mu s$ of data around the pulse peak to obtain 1\,MHz channelised Stokes I, Q, U and V spectra. Figure~\ref{fig:vela_uwl} shows the results for the linear and circular polarisation fractions over the 15 UWL observations. Across the ASKAP observing band (700-1800 MHz), $L/I$ and $V/I$ are relatively stable over the 4 year dataset. Based on these results, we developed empirical models of $L/I$ and $V/I$ using a third order polynomial and a hyperbolic tangent function respectively:

\begin{figure*}[t]
    \centering
    \includegraphics[width=\linewidth]{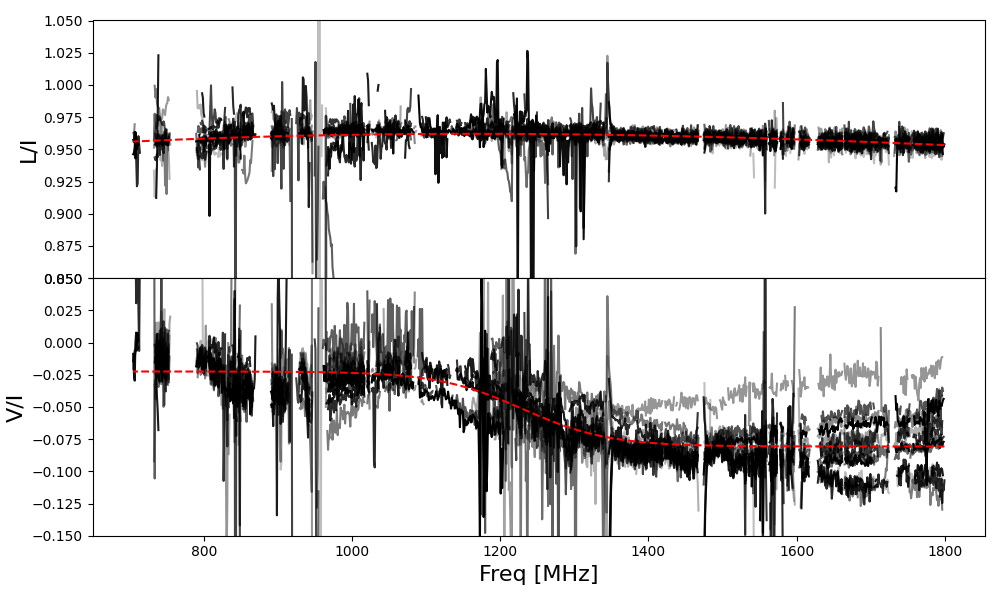}
    \caption{Summary of Vela observations from 26th December 2020 to 4th December 2024 zoomed using the UWL receiver. The spectra are zoomed in on the ASKAP observing band. Top Panel: The linear polarisation fraction from the UWL observations are shown in black; the different shades of black are used to differentiate the 15 observations. The red dashed line shows the best fit for L/I using a third order polynomial. Bottom Panel: The circular polarisation fraction. The red dashed line shows the best fit for V/I using a hyperbolic tangent function.}
    \label{fig:vela_uwl}
\end{figure*}

\begin{equation}
    \begin{split}
        &L_{\mathrm{o}} = 5.62\times10^{-12}f^{3} -4.51\times10^{-8}f^{2}+8.24\times10^{-5}f \\ & +9.18\times10^{-1} \\
        &V_{\mathrm{o}} = 2.91\times10^{-2}\tanh(-8.62\times10^{-3}(f + 5.16\times10^{-2})) \\
        & + 1.23\times10^{3},
    \end{split}
\end{equation}

\noindent where $\tanh$ is defined in \citet{tanh}. The difference between the two models is only a few percent; by default, we use the flat-spectrum model. 

We use FRB\,20230708A to demonstrate the new pipeline using Vela as the calibrator with a flat spectrum model; the details of the detection and offline processing are already published \citep{Dial2025}. Before deriving the calibration solutions we ensure the Stokes dynamic spectra of the calibrator are properly de-dispersed. The calibrator voltages are already de-dispersed coherently at this stage \citep{Scott2023}; for Vela a nominal DM of 67.97 pc cm$^{-3}$ is used. However, the DM of Vela is known to change by $\sim$0.2 pc cm$^{-3}$ yr$^{-1}$. Therefore, we perform a second stage of incoherent dedispersion using $\Delta$DM. A number of trial $\Delta$DMs are performed and the highest S/N is used as input to derive the calibration solutions.

\begin{figure}[t]
    \centering
    \includegraphics[width=0.9\linewidth]{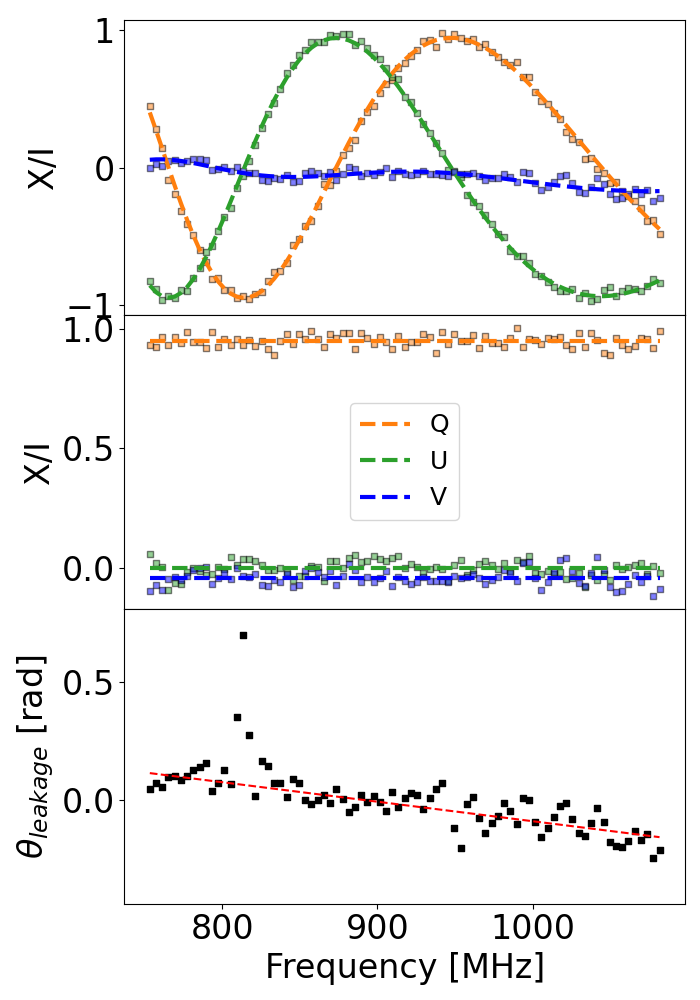}
    \caption{Top panel: The measures Stokes \textit{Q}, \textit{U} and \textit{V} spectra of Vela and the derived polarisation calibration models shown in the bold dashed lines. Middle panel: The Vela Stokes spectra after the polarisation corrections have been applied and the intrinsic Vela model shown in the bold dashed lines. Bottom panel: Best fit for the polarisation leakage.}
    \label{fig:VELA_fitting}
\end{figure}

\begin{figure}[t]
    \centering
    \includegraphics[width=\linewidth]{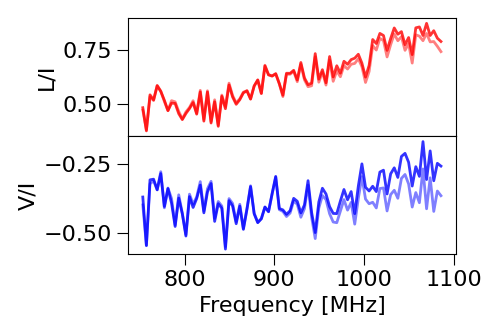}
    \caption{Stokes linear polarisation (top panel) and circular polarisation (bottom panel) for FRB20230708A with the applied polarisation calibration solutions (darker lines) and without (fainter lines). The spectra were obtained by integrating over the first bright component of the burst (see \cite{Dial2025}).}
    \label{fig:230708_polcal}
\end{figure}

Figure \ref{fig:VELA_fitting} shows the fitted calibration solutions using Vela. These solutions were applied to the voltages of FRB\,20230708A. We used the \texttt{ILEX}\footnote{https://github.com/tdial2000/ILEX} software package to process the voltages and obtain the Stokes spectra that are shown in Figure \ref{fig:230708_polcal}. Although the deviations in the observed Vela polarisation are on the order of a few \% when compared to the intrinsic model, the changes in \textit{L/I} and \textit{V/I} can be significant, which illustrates the importance of performing proper polarisation calibration. 

\begin{figure}
    \centering
    \includegraphics[width=\linewidth]{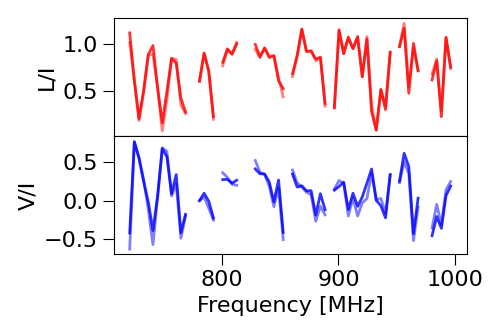}
    \caption{Stokes linear and circular polarisation fractions for FRB20240210A with a beam offset of 1.71 degrees.}
    \label{fig:240210_polcal}
\end{figure}

We investigated the effects of beam offset on these solutions. For offsets up to 0.5 degrees (which is the case for most FRBs), deviations from the Vela model remain within a few percent. However, for bursts detected at large offsets ($>$1$\degree$), beam morphology begins to drastically affect the measured polarisation. The current modelling cannot fully capture these effects as evidenced by the strong oscillations in \textit{L/I} and \textit{V/I} seen in Figure~\ref{fig:240210_polcal}.

\section{CELEBI performance and additional tools}\label{sec:tools}

\subsection{Structure maximisation - A SHRINE for CELEBI}\label{sec:shrine}

The observed DM of an FRB is important to accurately measure. Dedispersing the FRB signal is particularly key to properly investigate the temporal and polarised properties of the wide range of burst structures \citep[as demonstrated in][]{Scott2025}, as studies on FRB pulses are sensitive to the assumed value of DM used. There are two main approaches: a signal-to-noise (S/N) ratio maximising DM value, or one that maximises structure \citep[e.g.][]{Caleb2020, Platts2021}. \citet{Sutinjo2023} introduced a smoothing filter method and structure maximisation method (the vector norm or `2-norm', $\sqrt{\Sigma(d/dt)^{2}}$) which included a calculation of the uncertainty of the structure parameter. The approach used by \citet{Sutinjo2023} was in turn developed into a separate package and applied to derive the reported DM values for the CRAFT ICS (incoherent sum survey) catalogue \citep{Shannon2024}. These DM values were also used for the presented CRAFT HTR FRB pulses analysed by \citet{Scott2025}.

Due to the data processing workflow of CELEBI in localising the FRB from voltage data, extracting an HTR dataset using the localisation, and then refining the localisation based on improved DM and matched filtering before repeating the HTR step, this process was built into the CELEBI code directly. However, as this structure maximisation approach to the DM can be employed in FRB datasets outside of those obtained through CRAFT, the code for deriving structure maximising DM, as well as S/N maximised values of DM, has additionally been made publicly available. The SHRINE package (Structure maximisation of High-time Resolution Intensity profiles with No-nonsense Errors) can be found at \url{https://github.com/marcinglowacki/SHRINE}. 

%\subsection{ILEX}\label{sec:ilex}

%Github link, summary, some examples\\
%RM fitting\\
%polcal\\
%PA\\
%dynamic spectrum\\
%scattering\\
%period pulse search 

\subsection{Profile-based scientific workflow management}\label{sec:profile_workflow}
To facilitate the deployment of CELEBI on distributed super-computing infrastructure, the Nextflow workflow management system is employed~\citep{di2017nextflow}.
The CELEBI pipeline is specified using Nextflow as a directed acyclic graph of tasks, each of which is specified with a Python or Bash script.
Each task is associated with a list of input and output files, as well as a list of global, workflow-level parameters that the task depends on, such as the number of antennas or the observation duration.
Nextflow automatically handles the scheduling of CELEBI's tasks on the nodes of the super-computing infrastructure and the management of files transiting between tasks and nodes.

Experience gained from the first version of CELEBI revealed inefficiencies in Nextflow's resource management.
In particular, the application developer is responsible for specifying the resources reserved for each task, such as the number of CPU cores, the amount of memory reserved, or the duration of the allocated time slot.
These specifications were usually based on rough task requirement estimates, resulting in static over-provisioning of execution time by 94\% and memory footprints by 89\% for a typical CELEBI execution.
This over-provisioning is partly due to the configuration-dependent nature of the time and memory requirements of tasks, which causes developers to set large reservations that are compatible with any pipeline configuration.
Consequently, the latency of the pipeline on a shared super-computing infrastructure deteriorates as the allocation of unnecessarily large tasks is less likely to happen quickly.

To allocate resources more effectively for CELEBI's tasks, an analysis of prior execution traces can be used to build a configuration-dependent resource allocation model %, as described in~\cite{desnos202x}
(Desnos et al., in prep.)\footnote{\url{https://github.com/kdesnos/nextflow\_toys}}.
Each execution of the Nextflow pipeline generates execution traces containing resource monitoring data for each task. 
These execution traces are obtained from multiple runs of the input pipeline with different configurations. Configuration-dependent prediction models for memory and execution time are then constructed from these execution traces using statistical and machine learning methods, such as ANOVA (ANalysis Of VAriance) and symbolic regressions.
These models can then be used to generate new configurations for future pipeline executions, where the resource requirements for each task are based on these prediction models.
Using this approach with $3\sigma$ safety margins to avoid task failures due to insufficient allocated resources, over-provisioning of task execution time is reduced to 33\%, and memory over-provisioning to 23\%, with CELEBI.
On the shared OzSTAR infrastructure, these finer task granularities lead to a substantial reduction ($\sim$30-60\%) in the wall-time latency of the pipeline in uncontrolled conditions with a varying supercomputer workload, thanks to faster scheduling of smaller tasks (while also helping the OzSTAR queue to be scheduled more efficiently).

\subsection{Software container environment}\label{sec:container}

The previous version of CELEBI was dependent upon specific software dependencies and operating system (OS) of nodes on the OzSTAR (Optical \& X-ray Supercomputing for Theoretical Astronomical Research) supercomputing facility\footnote{\url{https://supercomputing.swin.edu.au/ozstar}}. In 2023, a change of the OS of OzSTAR to match a newer Ngarrgu Tindebeek (NT) machine was coupled with new system libraries and software modules. This system upgrade was incompatible with the original version of CELEBI. In order to address this, avoid issues from any other future changes in OS or system libraries, and ensure CELEBI could operate on other supercomputers, we opted to adapt CELEBI to a containerised system. Installation of the container henceforth is the only prerequisite with regard to software for CELEBI to be able to function. Special thanks to Astronomy Data and Computing Services (ADACS) is given for their help with this process, alongside error handling and  optimisation improvements to CELEBI (typical improvement in run-time by over a factor of two). 

The software container CELEBI now uses was developed through Docker, and can be loaded either locally or on a supercomputer with Docker installed. The container includes an installation of the Astronomical Image Processing System (AIPS), a previously difficult component to install and vital dependency of CELEBI. Other dependencies of CELEBI captured in the container include the correlator program Distributed FX \citep[DiFX;][]{Deller2007, Deller2011}, the Very Long Baseline Interferometry (VLBI) scheduling program SCHED, and Common Astronomy Software Applications \citep[CASA;][]{CASATeam2022}. The container is maintained and is publicly available on GitHub: \url{https://github.com/marcinglowacki/celebi-container}.

\section{Conclusion}\label{sec:conclusion}

The PINK update for CELEBI has improved both our astrometry precision of fast radio transient localisation, and the reliability of the polarisation measurement of such transient signals. New localisation improvements are particularly vital for lower S/N FRBs, expected to only be detectable with the improved CRACO detection system and limited to $>$arcsecond precision. Higher DM FRBs also particularly benefit from a matched filter imaging technique now implemented in order to confirm host association. FRB\,20251019A is an example of a transient detectable at relatively low S/N with the CRACO system as well as at high DM ($>1000$~pc\,cm$^{-3}$), where we significantly improve localisation precision through matched filter algorithms, as well as reliable astrometry correction of the RACS catalogues. While FRB\,20251019A currently remains hostless, other FRBs are expected to benefit from the PINK upgrades to CELEBI. Updates to our polarisation modelling meanwhile account for previously unappreciated issues with polarisation leakage, while leakage issues with edge and corner PAF beams have been identified.

We remind the reader that other upgrades to CELEBI, e.g. the use of deeper field images for astrometry correction, are planned. CELEBI is now fully portable as well with Docker software containerisation maintained on GitHub, enabling its use on other computing facilities and with future radio telescope systems. 

\begin{acknowledgement}

We thank the anonymous referee for their useful feedback which helped improve this paper. This work was supported by software support resources awarded under the Astronomy Data and Computing Services (ADACS) Merit Allocation Program. ADACS is funded from the Astronomy National Collaborative Research Infrastructure Strategy (NCRIS) allocation provided by the Australian Government and managed by Astronomy Australia Limited (AAL). This scientific work uses data obtained from Inyarrimanha Ilgari Bundara, the CSIRO Murchison Radio-astronomy Observatory. We acknowledge the Wajarri Yamaji People as the Traditional Owners and native title holders of the Observatory site. CSIRO’s ASKAP radio telescope is part of the Australia Telescope National Facility (https://ror.org/05qajvd42). Operation of ASKAP is funded by the Australian Government with support from the National Collaborative Research Infrastructure Strategy. ASKAP uses the resources of the Pawsey Supercomputing Research Centre. Establishment of ASKAP, Inyarrimanha Ilgari Bundara, the CSIRO Murchison Radioastronomy Observatory, and the Pawsey Supercomputing Research Centre are initiatives of the Australian Government, with support from the Government of Western Australia and the Science and Industry Endowment Fund. We also thank the MRO site staff. This work was performed on the OzSTAR national facility at Swinburne University of Technology. The OzSTAR program receives funding in part from the Astronomy National Collaborative Research Infrastructure Strategy (NCRIS) allocation provided by the Australian Government, and from the Victorian Higher Education State Investment Fund (VHESIF) provided by the Victorian Government. 
Some of the data presented herein were obtained at the W. M. Keck Observatory, which is operated as a scientific partnership among the California Institute of Technology, the University of California and the National Aeronautics and Space Administration. The Observatory was made possible by the generous financial support of the W. M. Keck Foundation.
The authors wish to recognise and acknowledge the very significant cultural role and reverence that the summit of Maunakea has always had within the indigenous Hawaiian community.  We are most fortunate to have the opportunity to conduct observations from this mountain.
This publication makes use of data obtained through Swinburne Keck program 2025B\_W007.

\end{acknowledgement}

\paragraph{Funding Statement}

MG is supported by the Australian Government through the Australian Research Council’s Discovery Projects funding scheme (DP210102103), and through UK STFC Grant ST/Y001117/1. MG acknowledges support from the Inter-University Institute for Data Intensive Astronomy (IDIA). IDIA is a partnership of the University of Cape Town, the University of Pretoria and the University of the Western Cape. For the purpose of open access, the author has applied a Creative Commons Attribution (CC BY) licence to any Author Accepted Manuscript version arising from this submission. AB acknowledges support through project CORTEX (NWA.1160.18.316) of the research programme NWA-ORC which is financed by the Dutch Research Council (NWO). A.C.G. and the Fong Group at Northwestern acknowledge support by the National Science Foundation under grant Nos. AST-1909358, AST-2206494, AST-2308182 and CAREER grant No. AST-2047919. This project has received funding from the European Union’s Horizon 2020 research and innovation programme under the Marie Skłodowska-Curie grant agreement No 873120. A.C.G acknowledges support from NSF grants AST-1911140, AST-1910471 and AST-2206490 as a member of the Fast and Fortunate for FRB Follow-up team. RLD is supported by the Australian Research Council through the Discovery Early Career Researcher Award (DECRA) Fellowship DE240100136 funded by the Australian Government. RMS acknowledges support through Australian Research Council Discovery Project DP220102305. 

%\paragraph{Competing Interests}

%A statement about any financial, professional, contractual or personal relationships or situations that could be perceived to impact the presentation of the work --- or `None' if none exist.

\paragraph{Data Availability Statement}

Data can be made available upon reasonable request to the CRAFT team. CELEBI, its software container, and SHRINE are publicly available through GitHub. 
%A statement about how to access data, code and other materials allowing users to understand, verify and replicate findings --- e.g. Replication data and code can be found in Harvard Dataverse: \verb+\url{https://doi.org/link}+.

%\endnote in some journals will behave like \footnote; and \printendnotes will not output anything. 
\printendnotes

\printbibliography

%\appendix

%\section{Example Appendix Section}

\end{document}